\documentclass[apj]{emulateapj}

\usepackage{amsmath}
\usepackage{xspace} 
\usepackage{graphicx}
\usepackage{txfonts}
\usepackage{rotate}
\usepackage{fancyheadings}

\newcommand{\un}[1]{~\hspace{-1pt}\ensuremath{\mathrm{#1}}}
\long\def\symbolfootnote[#1]#2{\begingroup\def\thefootnote{\fnsymbol{footnote}}\footnote[#1]{#2}\endgroup}

\newcommand{\as}{$^{\prime\prime}$\xspace}
\newcommand{\ergpersec}{\;erg\,s$^{-1}$\xspace}
\newcommand{\ergpercmsec}{\;erg\,cm$^{-2}$\,s$^{-1}$\xspace}
\newcommand{\percmsec}{cm$^{-2}$\,s$^{-1}$\xspace}

\newcommand{\integ}{{\it Integral}\xspace}
\newcommand{\xmm}{{\it XMM-Newton}\xspace}
\newcommand{\chandra}{{\it Chandra}\xspace}

\newcommand{\mos}{Mos\xspace}
\newcommand{\pn}{PN\xspace}
\newcommand{\epic}{Epic\xspace}

\newcommand{\sgra}{Sgr\,A*\xspace}
\newcommand{\cxo}{CXOGC\ J174540.0-290031\xspace}

\newcommand{\xrays}{X-rays\xspace}
\newcommand{\xray}{X-ray\xspace}

\newcommand{\bk}{\hspace{-6pt}}

\submitted{2005 January 27}
\received{2005 January 27}
\revised{2005 July 28}

\begin{document}

\title{Repeated X-ray Flaring Activity in Sagittarius A*}

\author{G. B\'elanger,\altaffilmark{1}\altaffilmark{2} A. Goldwurm,\altaffilmark{1}\altaffilmark{2}, F. Melia\altaffilmark{3}, P. Ferrando,\altaffilmark{1}\altaffilmark{2}, N. Grosso\altaffilmark{4}, D. Porquet\altaffilmark{5}, R. Warwick\altaffilmark{6}, F. Yusef-Zadeh\altaffilmark{7}}

\altaffiltext{1}{\scriptsize Service d'Astrophysique, DAPNIA/DSM/CEA, 91191 Gif-sur-Yvette, France; belanger@cea.fr}
\altaffiltext{2}{\scriptsize Unit\'e mixte de recherche Astroparticule et Cosmologie, 11 place Berthelot, 75005 Paris, France}
\altaffiltext{3}{\scriptsize Department of Physics and Steward Observatory, University of Arizona, Tucson, AZ 85721, USA}
\altaffiltext{4}{\scriptsize Laboratoire d'Astrophysique de Grenoble, Universit\'e Joseph-Fourier, 38041 Grenoble, France}
\altaffiltext{5}{\scriptsize Max-Planck-Institute f\"{u}r extraterrestrische Physik, Munich D-85741, Germany} 
\altaffiltext{6}{\scriptsize Department of Physics and Astronomy, University of Leicester, Leicester LE1 7RH, UK}
\altaffiltext{7}{\scriptsize Department of Physics and Astronomy, Northwestern University, Evanston, IL 60208}

\begin{abstract}
Investigating the spectral and temporal characteristics of the \xrays coming from 
Sagittarius A* (\sgra) is essential to our development of a more complete understanding 
of the emission mechanisms in this supermassive black hole located at the center of 
our Galaxy.  Several \xray flares with varying durations and spectral features have 
already been observed from this object.
Here we present the results of two long \xmm observations of the Galactic nucleus carried 
out in 2004, for a total exposure time of nearly 500\un{ks}.  During these observations we 
detected two flares from \sgra with peak 2--10\un{keV} luminosities about 40 times
($L_{\rm X}$\,$\sim$\,9\,$\times$\,$10^{34}$ \ergpersec) above the quiescent 
luminosity: one on 2004 March 31 and another on 2004 August 31. The first flare lasted 
about 2.5 ks and the second about 5\un{ks}. 
The combined fit on the \epic spectra yield photon indeces of about 1.5 and 1.9 for
the first and second flare respectively. This hard photon index strongly suggests
the presence of an important population of non-thermal electrons during the event 
and supports the view that the majority of flaring events tend to be hard and not very luminous.
\end{abstract}

\keywords{black hole physics --- Galaxy: center --- 
			Galaxy: nucleus --- \xrays: observations ---
			stars: neutron --- \xrays: binaries}

\section{Introduction}
\label{s:intro}

Sagittarius A*, the black hole at the Galactic center,
is a source of fascination and curiosity for many in the
astrophysical community.  There are several reasons for this, including
the fact that \sgra provides us with the most compelling evidence to date 
for the existence of supermassive black holes in the universe and is the 
closest such object (Sch\"odel et al.\ 2003; Ghez et al.\ 2003).  Its 
relative proximity allows us to investigate \sgra's radiative characteristics, 
as well as those of its nearby environment, with great detail and excellent
spatial and spectral resolution in all of the accessible wavebands.
In addition, unlike more typical active and bright galactic nuclei,
\sgra is very dim and shines at less than $10^{-9}$ \citep{c:melia01}
times the Eddington luminosity for an object of its mass, now thought to be 
$M$\,$\approx$\,3.4\,$\times$\,$10^{6}$\,$\rm{M}_{\odot}$ (Sch\"odel et al.\ 2003).
This faintness is itself intriguing and raises many more questions about 
the conditions that exist in the environment surrounding the massive black hole.  

And yet, this faintness may actually be a blessing in disguise for probing the 
innermost regions of this object, for it points to an optically thin envelope that is almost 
transparent to high-energy IR and \xray photons produced during transient events 
manifested within a mere handful of Schwarzschild radii ($r_S$\;$\equiv$\;$2GM/c^2$, 
or roughly 9\,$\times$\,$10^{11}$\un{cm}) of the event horizon.
Thus, the Galactic centre's supermassive black hole provides a wonderful laboratory 
for the study of accretion and related phenomena, including the formation and 
evolution of accretion disks, the causes and effects of magnetic reconnection,
the emission characteristics from the acceleration of charged particles in the 
strong magnetic and gravitational fields, the relationship between
radiation at different wavelengths, and the processes that give rise to these 
(see Melia \& Falcke 2001 for a review).

In early 2001, the \chandra \xray Observatory detected what we now consider to be 
the first \xray counterpart (Baganoff et al. 2003) to the radio source \sgra.  The 
quiescent state of this source was characterized by a power-law photon index 
$\Gamma$\,=\,$2.7^{+1.3}_{-0.9}$ and a derived 2--10\un{keV} luminosity of
$L_{\rm X}$\,=\,(2.2\,$\pm$\,0.4)\,$\times$\,$10^{33}$\ergpersec.
It is thought that the same population of electrons that produce the synchrotron 
peak at sub-mm wavelengths give rise to this quiescent component through
synchrotron self-compton (SSC) (Liu \& Melia 2001; Markoff et al. 2001).

In the fall of the same year, \chandra detected the first \xray flare 
from this source \citep{c:baganoff01}.  This flare lasted about 10\un{ks},
had a harder photon index of $\Gamma$\,=\,1.3\,$\pm$\,0.6,
and reached a peak luminosity of $L_{\rm X}$\,=\,(1.0\,$\pm$\,0.1)\,$\times$\,$10^{35}$ 
\ergpersec (or factor 45 above the quiescent level).  To this day,
two additional bright flares from the direction of \sgra, with varying time scales 
and spectral features, have been detected and reported.  Goldwurm et al.\ (2003a) 
reported the detection of the rising portion of a flare from \sgra 
over some 900\un{s}.  They found a flare photon index of $\Gamma$\,=\,0.7\,$\pm$\,0.6, 
compatible with the \chandra flare photon index within the errors, and they derived 
a peak luminosity of $L_{\rm X}$\,=\,(5.4\,$\pm$\,1.0)\,$\times$\,$10^{34}$\ergpersec
(a factor 25 above quiescence).  The most recent report of a flare from \sgra is that 
presented by Porquet et al. (2003).  This flare was quite different from the two previous 
events, for it lasted roughly 2.7\un{ks}, had a soft spectral index ($\Gamma$\,=\,2.5\,$\pm$\,0.3), 
and reached a surprisingly high peak 2--10\un{keV} luminosity of
$L_{\rm X}$\,=\,(3.6\,$\pm$\,0.4)\,$\times$\,$10^{35}$\ergpersec. This corresponds
to an unprecedented factor 160 above the source's quiescent luminosity.

\xray flaring activity in \sgra could be induced by a sudden enhancement of 
accretion \citep{c:liu02}, from a swift acceleration of electrons 
in a magnetic flare near the black hole (Liu, Petrosian, and Melia 2004), within 
a jet \citep{c:markoff01}, or perhaps through some other nonthermal process within
a radiatively inefficient accretion flow (Yuan, Quataert, \& Narayan 2003 and 2004).
Furthermore, short time scale \xray and IR flares could be much more difficult to detect 
if the luminosity of \sgra were even slightly larger in which case the thermal SSC
 emission would possibly dominate over the non-thermal emission
(Yuan, Quataert, \& Narayan 2004).  The recent detection of \sgra in the near-IR band 
in both the quiescent and flaring states \citep{c:genzel03},
and the more recent detection of a correlated increase in near-IR and 
\xray flux (Eckart et al.\ 2004), provide further important constraints 
on building a comprehensive model of this source.  Although relatively few instances 
of detected flares from \sgra have been reported, 
there appears to be two ``types'' of flares---the soft and hard---and a model that 
can naturally explain both of these types of events is presented in
Liu, Petrosian, \& Melia (2004).

We here present the results of two long \xmm observations of \sgra and the Galactic 
center in which we detected two flares that rose to a factor of 40 above the quiescent 
luminosity from the direction of the central black hole, and several other, less 
significant events that could also be coming from \sgra.
Both of these observations were conducted concurrently with \integ in order to 
investigate possible links in the manifestation of a sudden enhanced accretion, 
magnetic reconnection, or other types of events in the hard \xray (2--12\un{keV}) 
and soft gamma-ray (20--120\un{keV}) domains. 
Furthermore, a set of other observations of \sgra covering the radio, 
sub-millimeter and infrared frequencies, coordinated with the 2004 \xmm large project,
were also planned and partly performed, and two new transient radio sources were 
discovered with the VLA \cite{c:bower05}.
Results from the \integ observations \citep{c:belanger05a}
and from the overall multiwavelength campaing will be reported elsewhere.
The two main flares, which we will refer to as the factor-40 flares hereafter, 
occurred on time scales between 2500--5000\un{s} 
and have similar spectral characteristics.
These may be added to the statistics of detected flares from 
\sgra and add weight to the interpretation that there are
indeed two types of flares, bright and soft or not-so-bright and hard;
and that the majority of these events are of the latter kind. 
A larger sample of such events is essential for constructring a more robust 
classification---and thus identification---of the causes for this flaring activity in \sgra.  
Furthermore, a brightening of the region around \sgra by a factor of $\sim$\,2 in 
the 2--10\un{keV} band with respect to all previous \xmm observations of the 
Galactic center is evident in both datasets.
The astrometry-corrected fitted centroid of this emission
coincides with the position of a bright transient lying about 3\as south of \sgra 
and discovered by \chandra in the summer of 2004 \citep{c:muno05a}.
Here we will only briefly comment on the emission characteristics of this new source,
\cxo, during the \xmm observations and refer the interested 
reader to Porquet et al. (2005) and Muno et al. (2005b) for further details.

The structure of this paper is as follows: In \textsection\ref{s:obs} we describe 
the observations and the analysis methods used to reduce the data.  
In \textsection\ref{s:results}, we present the results of the analysis, 
and in \textsection\ref{s:discussion}, we discuss their possible implications 
and interpretations.

\section{Observations and Methods of Analysis}
\label{s:obs}

\begin{deluxetable}{lccc}
\tabletypesize{\footnotesize}
\tablewidth{0pt}
\tablecolumns{4}
\tablecaption{Observation Log}
\tablehead{
ObsID      & \colhead{Obs Start time} & \colhead{Obs End time} & \colhead{Duration} \\
\colhead{} &    \colhead{(UT)}        &    \colhead{(UT)}      & \colhead{(ks)}
}
\startdata
788		& 2004-03-28T14:57:08	&  2004-03-30T04:44:00 & 112.6 \\
789		& 2004-03-30T14:46:36	&  2004-04-01T04:35:49 & 123.4 \\
866		& 2004-08-31T03:12:01	&  2004-09-01T16:45:58 & 130.1 \\
867		& 2004-09-02T03:01:39	&  2004-09-03T16:35:35 & 131.4
\enddata
\label{t:obsLog}
\end{deluxetable}

The \xmm satellite observed the Galactic center as part of a large
project during revolutions 788 and 789, between 2004 March 28 and April 1, 
and then during revolutions 866 and 867, between 2004 August 31 and September 2 
for a total exposure time of $\sim$\,490\un{ks}, during which the Epic \mos and \pn 
cameras were in {\it PrimeFullWindow} and 
{\it PrimeFullWindowExtended} modes, respectively, during epoch\,1,
and in {\it PrimeFullWindow} during epoch\,2.
The log of the observations is presented in Table\,\ref{t:obsLog}.
We will use epoch\,1 to refer to the observation period spanning
ObsID 788 and 789, and epoch\,2 for the one spanning ObsID 866 and 867.

We generated event lists for the \mos1, \mos2, and \pn cameras using the 
{\bf\tt emchain} and {\bf\tt epchain} tasks of the \xmm Science Analysis System v\,6.1.0.
These were subsequently filtered and used to construct images in two energy bands: 
0.5--2\un{keV}, and 2--10\un{keV}.
The filter on the event pattern in imaging mode,
({\tt PATTERN}\,\textless\,=\,12 for \mos, and {\tt PATTERN}\,\textless\,=\,4 for \pn),
ensures that only events created by \xrays and free of cosmic ray contamination 
are selected. Artifacts from the calibrated and concatenated datasets, 
as well as events near CCD gaps or bad pixels are rejected by setting 
({\tt FLAG==0}) as a selection criterion.
A further selection on the maximum count rate in the 10--12\un{keV} range
(18\un{cts/s}) for \mos and 12--14\un{keV} range (22\un{cts/s}) for \pn was 
applied to exclude all periods of increased charged particle flaring activity. 
This stringent selection criterion was only applied to the image construction.

The heavy absorption in the direction of the Galactic center, prevents photons 
from this region having energies below 1.5--2\un{keV} from reaching us.
For this reason, we used the 0.5--2\un{keV} images of foreground stars
to identify counterparts for astrometric corrections, and the 2--10\un{keV} image
to determine the flare centroid.

\subsection{Astrometry}
\label{s:astrom}

We identified three {\it Tycho-2} and two \chandra sources with positional 
uncertainties of \mbox{0\as\bk.025} and \mbox{0\as\bk.16}, respectively\,%
\footnote{
The astrometric sources are: 
Tycho-2\ 6840-20-1, 6840-666-1, 6840-590-1,
CXOGC\ J174545.2-285828 and CXOGC\ J174607.5-285951. 
Tycho-2 positions were taken from 
{\tt http://www.astro.ku.dk/\textasciitilde\/cf/CD/data/catalog.dat}
and astrometric precision from Hog et al. (2000).
\chandra positions are from Muno et al. (2003) and errors
are from Baganoff et al. (2003a)
}.
Of these calibration sources, three were in the central \mos CCD 
and two just outside of it.
This is an essential point in the astrometric corrections
because the positional uncertainty of a \mos CCD relative
to another is $\sim$ \mbox{1\as\bk.5} and this therefore determines the
minimum systematic positional uncertainty over the whole field of view. 
However, if we have at least three calibration sources in the central CCD, 
then we can reduce this systematic uncertainty for sources in that CCD.

The astrometric correction was done by first running {\tt edetect\_chain} using
the 0.5--2\un{keV} images to get a list of the detected sources with their fitted 
position and associated statistical uncertainty.
Then, using {\tt eposcorr} that optimizes the correlation between the positions of
the calibration sources and their \xray counterparts allowing for a displacement 
in R.A. and Decl. as well as rotation, we found the boresight corrections
for the three \epic instruments. The \mos cameras have smaller pixels, and 
therefore a finer angular resolution than the \pn instrument.
We therefore used the \mos images for our astrometric study of the obervations 
during which the flares occured, namely ObsID\,789 and 866. 
The results of this study are listed in Table\,\ref{t:astrom} where we give the 
offsets and root mean square (rms) dispersion in the positions of the astrometric 
sources with respect to their reference positions before and after correction.

Although we performed this procedure for each individual camera as well
as for the merged \mos event list we used the \mos2 results for our analysis
of the first flare (ObsID\,789) and the \mos1 for the analysis of the second 
(ObsID\,866) since these had the smallest dispersion after boresight corrections.
We found that the merged \mos image systematically yielded smaller offsets  
in both coordinates and that the rms dispersion remained in the range
0\as\bk.9 to 1\as\bk.2 before and after boresight corrections were applied.
This behaviour is expected since the astrometric precision derived from calibrations
and that rely upon the knowledge of the position of one camera with respect to the 
other is stated as 1\as\bk.2 \citep{c:kirsch05}. 
So even though we can generally expect smaller offsets and dispersions before 
corrections when using the merged \mos image as this naturally averages the 
astrometry of both \mos cameras, the minimum dispersion after corrections will 
always be around 1\as\bk.2. 
Unusually large dispersions can be caused by the ``splitting'' of one or more of the 
astrometric sources by a column of bad pixels as was seen in \mos2 during ObsID\,866.
Astrometric corrections maximize our ability to locate sources and to distinguish
a flaring source from its closest known neighbour as is our intention here. 

We found that the change in position due to corrections of up to 2.7\as in a given 
coordinate has negligible effects on the light curve and spectra that are 
constructed by integrating over a 10\as radius around the source. 
This is primarily due to the shape of the instrument point spread function 
which is characterized by a very narrow peak and broad wings. 
As long as the peak is included in the extraction region then variations due to a 
shifted centroid are completely negligible. 
Each individual event list is treated separately to ensure proper 
good-time-interval corrections.

\begin{deluxetable}{lcc}
\tabletypesize{\footnotesize}
\tablewidth{0pt}
\tablecolumns{3}
\tablecaption{Astrometric Corrections}
\tablehead{ \colhead{} 	& \colhead{ObsID\,789} 	& \colhead{ObsID\,866} 	}
\startdata
R.A. offset	\dotfill		&\hfill	$-2.67\pm0.60$	&\hfill	$0.88\pm0.61$	\\
Decl. offset \dotfill		&\hfill	$0.35\pm0.56$	&\hfill	$0.35\pm0.46$	\\
Rotation \dotfill		&\hfill	$-0.20\pm0.30$	&\hfill	$-0.05\pm0.25$ 	\\
RMS before correction \dotfill 	&	$3.22$		&	$1.40$ 		\\
RMS after correction \dotfill 	&	$0.59$		&	$0.88$	

\enddata
\tablecomments{\scriptsize 
All quantities are given in arc seconds. The column ObsID\,789 lists the parameter 
values for \mos2 and ObsID\,866 for \mos1. No significant rotational offset was 
detected in either case and therefore none was applied in the analysis.
}

\label{t:astrom}
\end{deluxetable}

\subsection{Light curve construction}
\label{s:lightcurves}

The source extraction region is defined as a circular area of radius 10\as
centrered on \sgra's radio position:
(J2000.0) R.A. = $\mathrm{17^{h}45^{m}40^{s}\hspace{-2pt}.0383}$, 
Decl. = $-29^{\circ}00^{\prime}28^{\prime\prime}\hspace{-3pt}.069$
\citep{c:yusef-zadeh99}.
The light curves for each observation were made from the event
files of the \epic instruments; \mos1, \mos2 and \pn, using 
the times of the first and last events overall to determine the bounds
used to define the temporal bins.
We constructed a rate set for each event file, calculated the effective bin time 
defined as the sum of the good-time-intervals (GTIs) within a bin,
applied the GTI corrections by multiplying each rate by the ratio
of nominal bin time to effective bin time, and finally summed these 
corrected rates for all temporally coincident bins.

To correct for background fluctuations even though
they are generally assumed to be negligible on a source 
extraction zone as small as 10\as,
we extracted a background light curve from an annular region 
with inner and outer radii of 10 and 500 arcsec around \sgra. 
These background rates were also GTI-corrected and then rescaled
to the area of the source extraction zone before being
subtracted from the source rates.
(The maximum background count rate we found for any given bin
was of the order of 10\% of the source count rate).

This procedure yielded the background subtracted, GTI-corrected light curves.
The GTI correction was particularly important for the data
of ObsID\,789 during which high levels of 
background solar flare particles caused the saturation
of buffers and thus the loss of data in the \pn camera by
switching it to counting mode for short time periods and so 
generating several short GTIs. In such instances, the count rate must
be estimated on longer time scales and thus larger binning.

For all light curves, we excluded the bins that correspond to time periods
that do not have simultaneous coverage by all three instruments, 
and those with very high count rates seen in only one of the instruments;
identified by looking at the ratio of the \mos to \pn count rates.
All light curves presented in this paper are combined \epic
light curves (\mos1\,+\,\mos2\,+\,\pn).

\subsection{Spectral extraction}

The source and background spectra for each flare were extracted over a circular 
region with a radius of 10\as centered on \sgra, and in each case, 
the backgrounds were integrated over about 50\un{ks} during ``quiescent'' 
periods --- excluding flares and eclipses --- as done by Goldwurm et al.\ 2003 
and Porquet et al.\ 2003.
This ensures that the spectrum is composed of almost only flare photons
such that the contribution from the underlying nearby emission is negligible. 
The time windows over which each flare spectrum was extracted were
optimized for maximum signal to noise ratio and are
[197161000:197163000] for the 2004 March 31 event and 
[210335000:210337500] for the 2004 August 31 flare, given as \xmm time 
in units of seconds (see Figures\,\ref{f:lc_789-flare} and \ref{f:lc_866-flare}). 
These intervals correspond to 2000 and 2500\un{s} respectively and
do not cover the entire event.

Since the source photons represent between about 35 and 50\% of the total 
counts, and that these are in all cases quite low in number,
we used the C-statistic to fit and derive the model parameters. 
In the context of this statistic, the source spectra do not have to be rebinned. 
However, the method requires a moderately well defined background with 
at least 5--8 counts per bin and we therefore grouped these using 
{\bf \tt grppha} such that each bin contains a minimum of 10 counts.
To ensure coherence between the source and background spectra, 
we applied exactly the same grouping to the source spectra.
Regrouping in larger bins, with 20 counts per bin for example,
gives parameter values that are statiscally consistent with the ones
derived from the ungrouped data set.

The redistribution and ancillary response matrices were generated for each 
source spectrum using the {\bf \tt rmfgen} and {\bf \tt arfgen} SAS tasks.

\section{Results}
\label{s:results}

\begin{figure}[ht]
\includegraphics[angle=-90,scale=0.35]{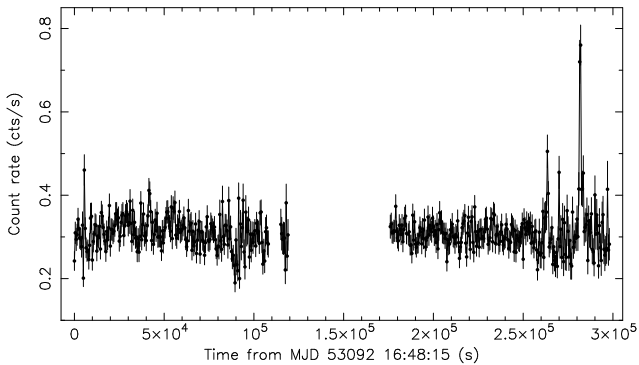}
\caption{\footnotesize 
	Light curve in the 2--10\un{keV} energy range for a circular region of radius 10\as 
	centered on \sgra binned in 500 second intervals for the epoch\,1 observations.
	MJD 53092 corresponds to 2004 March 28.
        \label{f:lc_788-789}}
\end{figure}

\begin{figure}[ht]
\includegraphics[scale=0.35,angle=-90]{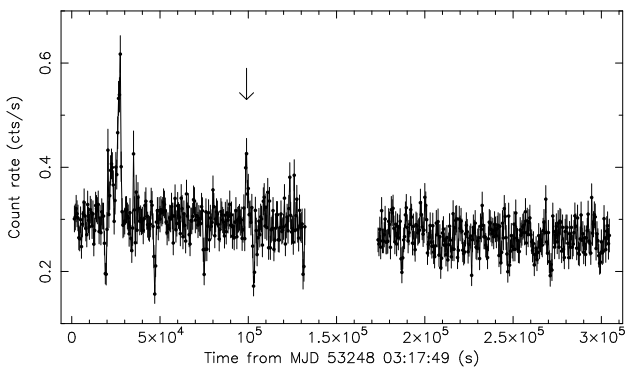}
\caption{\footnotesize
	Light curve in the 2--10\un{keV} range for a circular region of radius 10\as 
	centered on \sgra binned in 500 second intervals for the epoch\,2 observations.
	The arrow marks an isolated peak with statistical sigificance of about 5$\sigma$.
	MJD 53248 corresponds to 2004 August 31.
        \label{f:lc_866-867}}
\end{figure}

Figures\,\ref{f:lc_788-789} and \ref{f:lc_866-867} show the epoch\,1 and
epoch\,2 light curves respectively. Two large flares and a few smaller ones are 
apparent in the second portion of Figure\,\ref{f:lc_788-789} and the first 
portion of Figure\,\ref{f:lc_866-867}. More detailed views of the flaring periods are 
shown in Figures\,\ref{f:lc_789-flare} and \ref{f:lc_866-flare}. 

During epoch\,1, the flaring activity occurs towards the end the observation 
where a large event is preceeded by two smaller ones.
In epoch\,2, the large flare and its precursor occur near the start of the first observation.
In the latter, the main event is closely followed by a small peak, 
and another statistically significant flare closer to the end of ObsID\,866
(marked by an arrow in Figure\,\ref{f:lc_866-867}).
Two small peaks at the very end of the same observation have significances
just above 3$\sigma$.
In addition, there is a distinct periodic eclipse clearly seen over more than four 
complete cycles in the course of ObsID\,866 and also present in ObsID\,867
even though it is not as evident.

In the next section, we present a detailed survey of the results obtained for the 
most significant flares, followed by an analysis of the temporal
characteristics of the light curves.
We show that the two factor-40 flares are coming from 
the direction of \sgra and not from the transient binary system
\cxo to which the eclipses can be attributed (see Porquet et al.\ 2005). 

\subsection{Flare Analysis}

The position of the flaring source was determined by running 
{\tt \bf edetect\_chain} on an image of the central region constructed 
by selecting events in a temporal window corresponding to the duration of the flare,
and applying the astrometric corrections to the fitted position of the most significant detection.
The resulting flaring source position for the 2004 March 31 event was found to be:
R.A. = $\mathrm{17^{h}45^{m}40^{s}\hspace{-2pt}.08}$, 
decl. = $-29^{\circ}00^{\prime}28^{\prime\prime}\hspace{-3pt}.48$, J2000
with an uncertainty in the position of 1\as\bk.12, 
calculated by combining the statistical uncertainty on the fitted position (0\as\bk.44),
the boresight correction (0\as\bk.82), 
and the rms dispersion in the astrometric sources after correction (0\as\bk.62).
This position is at 0\as\bk.66 from the radio position of \sgra
and 2\as\bk.58 from the new transient \cxo,
the closest known \xray source to the central black hole.
This permits us to unambiguously associate the flaring source with \sgra.

For the 2004 August 31 flare, the position of the flaring source was found to be:
R.A. = $\mathrm{17^{h}45^{m}40^{s}\hspace{-2pt}.14}$, 
decl. = $-29^{\circ}00^{\prime}28^{\prime\prime}\hspace{-3pt}.40$, J2000
with a positional uncertainty of 1\as\bk.31.
This is 0\as\bk.52 from \sgra and 2\as\bk.98 from \cxo, therefore the association 
of this flaring source with the central black hole is once again unambiguous.
The centroid of the 2--10\un{keV} emission in the non-flare period 
of ObsID\,866 is located just 0\as\bk.71 from \cxo but 2\as\bk.38 from \sgra.
This strongly suggests that this source does indeed contribute a large portion of 
the \xray flux in this band but that it is without a doubt not the flaring source.

\begin{figure*}[ht]
\epsscale{1.0}
\begin{center}
\includegraphics[scale=0.7, angle=0]{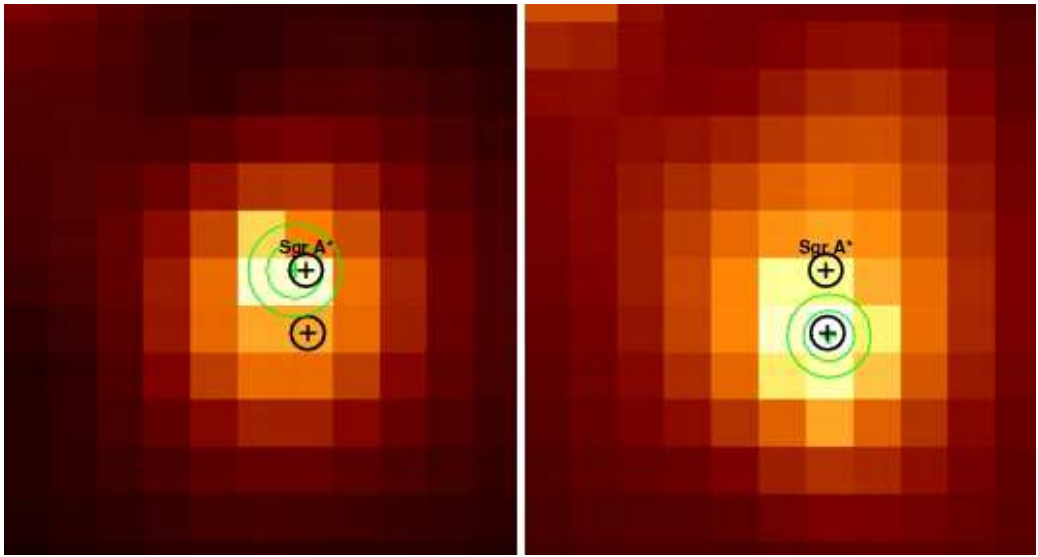}
\end{center}
\caption{\footnotesize 
	Smoothed \mos2 image of the Agust 31 flare with an exposure time of 2000\un{s} 
	(left panel) and average non-flare image with an exposure of about 50\un{ks} (right).
	The boresight corrected position of \sgra is labeled and that of \cxo is marked
	by a cross. The black circles indicate the 1$\sigma$ uncertainty on the position
	due to the astrometric correction. The green cross and circles mark the fitted
	emission centroid and the associated uncertainty at the 68 and 90\% confidence
	levels.
        \label{f:flare}}
\end{figure*}

\begin{figure}[ht]
\epsscale{1.0}
\includegraphics[scale=0.35,angle=-90]{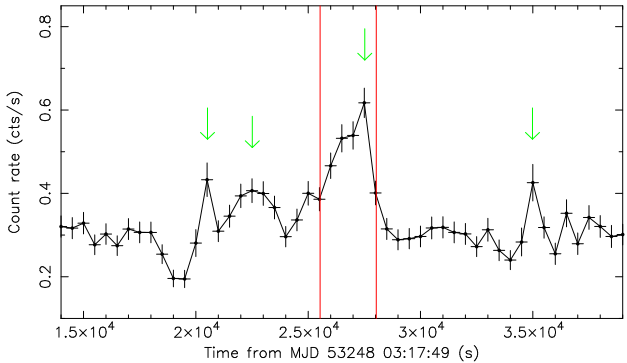}
\caption{\footnotesize Light curve of the 2004 August 31 flare binned in 500\un{s} intervals. 
	Green arrows point out local peaks with respective significances of about 
	4, 5, 10 and 4$\sigma$.
	We show the base level of the light curve before and after the flare for reference. 
	The drop in flux at the base of the precursor corresponds to the first of five 
	eclipses in the binary system \cxo detected during ObsID\,866.
	MJD 53248 corresponds to 2004 August 31.
	\label{f:lc_866-flare}}
\end{figure}

As an example, we show in the left hand side panel of Figure\,\ref{f:flare}, 
an image of the Galactic center during the August 31 flare composed of events 
selected from the same temporal window as the one used to build the flare 
spectrum (see Figure\,\ref{f:lc_866-flare}).
In black, the position of \sgra is crossed and labeled, that of \cxo is simply 
marked by a cross for clarity, and the circles show the uncertainty on the 
position after boresight corrections. 
In green, the cross marks the fitted position of the brightest source detected in the 
flare image, and the circle indicate the 68 and 90\% confidence regions derived 
by combining the statistical error on the fit (0\as\bk.60), 
the error from the boresight correction (0\as\bk.77), 
and the rms (0\as\bk.88) that we take as a systematic uncertainty,
On the right hand side, we see the non-flare image of the same region 
constructed by selecting all event used to build the background spectrum. 
The labeling scheme is the same as in the left panel.
Each pixel is 1\as\bk.1 in size matching the sky projected size of the 
camera's physical pixels. 
A wavelet filter was applied to smooth the image for presentation purposes only.

It is evident that the centroid of the emission during the flare is significantly
shifted towards \sgra and that the transient source \cxo can be excluded as the
flaring source at the 90\% confidence level.
In the non-flare period, the emission is very closely centered on the transient 
binary --- the central black hole's closest known, hard \xray emitting 
neighboor --- and in this case, \sgra can be excluded at more than the 90\% confidence  
level as the source of this emission. 
The error circles on the fitted centroid of the non-flare image are smaller because 
the statistical uncertainty on the fit is 0\as\bk.22 compared with 0\as\bk.60 for 
the flaring source. Combining this value with the same boresight correction 
uncertainty and rms as was used in the analysis of the flare image yields a total 
uncertainty of 1\as\bk.19.

\begin{figure}[ht]
\epsscale{1.0}
\includegraphics[scale=0.35,angle=-90]{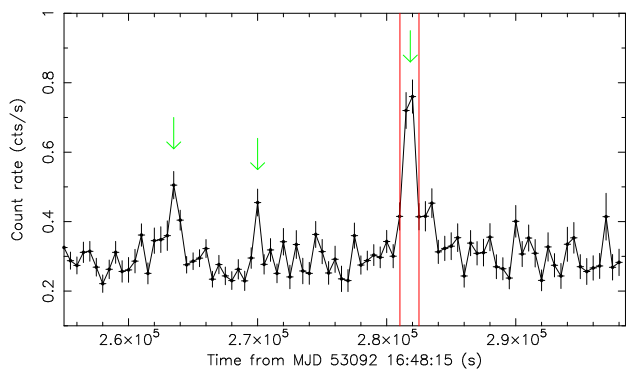}
\caption{\footnotesize 
	Zoom on the flare of 2004 March 31 with time bins of 500\un{s} 
	where the first factor-40 flare occured. Its significance is about 10$\sigma$. 
	The red lines delineate the time window used in the spectral extraction.
 	Two other smaller flares are marked by the green arrows and have
	significances of abotu 5 and 4$\sigma$ respectively.
	MJD 53092 corresponds to 2004 March 28.
	\label{f:lc_789-flare}}
\end{figure}

\begin{figure}[ht]
\epsscale{1.0}
\includegraphics[scale=0.35,angle=-90]{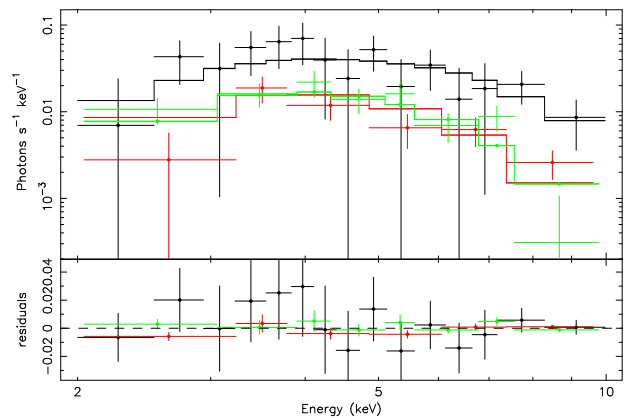}
\caption{\footnotesize Epic spectra during the \sgra flare of 2004 March 31. 
	The spectra are shown with the best fit absorbed power-law model
	and the corresponding residuals. The fit is done in the range 2--10\un{keV}.
	The \pn spectrum is shown in black and the \mos\,1 and \mos\,2
	are in red and green respectively.
        \label{f:789_flareSpecs}}
\end{figure}

The first flaring period occurred on 2004 March 31 (Figure\,\ref{f:lc_789-flare})
and contains one of the two factor-40 flares detected over the course of 
all four pointings, preceded by two smaller ones.
The most prominent flare took place around 23:05 and peaked at 0.76\un{cts/s} 
estimated on the basis of 500\un{s} time bins. 
This event lasted $\sim$\,2.5\un{ks} from rise to fall.
The \pn (black), \mos1 (red) and \mos2 (green) spectra for this flare
are shown in figure\,\ref{f:789_flareSpecs}.

We fitted each spectrum individually and also simultaneously.
The low statistics limits the reliability of the \mos results and for this reason we
present the best fit parameters for the \pn data as well as for the combined data set
in Table\,\ref{t:specParams}.
Three models were tested: absorbed power-law, black body and bremsstrahlung.
All give satisfactory fits.
We used the {\tt pegpwrlw} model in XSpec v\,11.3.1 in which the 
the total unabsorbed flux over the range of the fit is used as the normalization.
This allows the photon index and normalization to be fit as independent parameters
in the model and to derive the uncertainty on the flux directly from this normalization. 

For the pegged power-law model, the best fit values of the combined 
data set are a photon index of $\Gamma$\,=\,1.5\,$\pm$\,0.5 with a column density of 
$N_{\rm{H}}$\,=\,(8.3\,$\pm$\,2.5)\,$\times$\,$10^{22}$\un{cm^{-2}}.
The average unabsorbed flux during the flare was found to be 
(6.35\,$\pm$\,0.45)\,$\times$\,$10^{-12}$\ergpercmsec. 
Since the count rate at the peak is about twice the average,
we estimate that the maximum flux is also about twice the average and so the peak 
2--10\un{keV} luminosity is around 10\,$\times$\,$10^{34}$\ergpersec 
at a distance of 8\un{kpc}; a factor of 45 above 
quiescence\,\footnote{We refer to both events as factor-40 flares for simplicity.}.

The black body was best fit with a temperature of $1.9 \pm 0.3$\un{keV} and 
absorbing column of $(4.1 \pm 1.7) \times 10^{22} \un{cm^{-2}}$.
Finally, fitting a bremsstrahlung model resulted in an absorbtion column
of $7.8\pm1.9 \times 10^{22} \un{cm^{-2}}$ --- similar to that of the power-law --- and 
a temperature around 35\un{keV}. 
The error range on the temperature, however, 
is huge and therefore not constraining at all.
The large upper bounds in the error of the bremsstrahlung temperature probably 
reflect the fact that there is no statistically significant break in the spectrum.
The best fit parameter values, fluxes and luminosities for the power-law and
black-body models are given in Table\,\ref{t:specParams}.

The respective occurrence times, peaks and durations 
of the two small flares that preceded it are:
17:59, 0.51\,$\pm$\,0.04\un{cts/s} ($\sim$\,5$\sigma$) and 1500\un{s} for the first, 
and 19:48, 0.45\,$\pm$\,0.04\un{cts/s} ($\sim$\,4$\sigma$) and 1000\un{s} for the second.
These two events are marked by green arrows in Figure\,\ref{f:lc_789-flare}
Unfortunately, the short duration and lower signal-to-noise
ratio of these events prevents us from doing their spectral analysis.

\begin{figure}[ht]
\epsscale{1.0}
\includegraphics[scale=0.35,angle=-90]{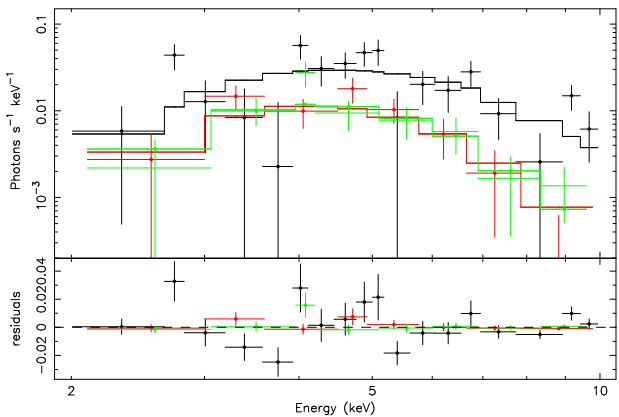}
\caption{\footnotesize  Epic spectra during the \sgra flare of
	2004 August 31 fitted in the range 2--10\un{keV}. 
	The colours are the same as in Figure\,\ref{f:789_flareSpecs}.
        \label{f:866_flareSpecs}}
\end{figure}

The second major flare was on 2004 August 31 and is shown
in detail in Figure\,\ref{f:lc_866-flare}.
The large flare is preceded very closely by what we will call a double-peaked precursor
that lasted about the same time as the flare itself.
Furthermore, we see a $\sim$\,30\% change in flux over about 900\un{s} or 15\un{min}
between the two peaks of the precursor which constrains the emitting region to less 
than 2 AU at a distance of 8\un{kpc}.
The first peak of the precursor occurred at around 9:00 and the flare that followed rose to 
its maximum at about 11:05 with a count rate of 0.62\,$\pm$\,0.03\un{cts/s} for 
500\un{s} bins.
Both the precursor and the flare lasted $\sim$\,5000\un{s} and so the whole flaring period
had a duration of about 10\,000\un{s} from rise to fall.
The narrow peak following the flare occurred at 13:01 and
the last flare occurred the next day, 2004 September 1, at 6:48.
Figure\,\ref{f:866_flareSpecs} shows the \pn (black), \mos\,1 (red), \mos\,2 (green) 
spectra of the flaring event with the best fit absorbed power-law model and residuals.

This flare was modeled in the same way as was done for the March 31 flare and 
the detailed spectral fitting results are listed in Table\,\ref{t:specParams}. 
The combined spectrum was best fit with an absorbed power-law of photon index
$\Gamma$\,=\,1.9\,$\pm$\,0.5 and column density 
$N_{\rm{H}}$\,=\,(12.5\,$\pm$\,3.4)\,$\times$\,$10^{22}$\un{cm^{-2}}.
For the black body model, the temperature was 1.8\un{keV} and the
absorbing column density (7.1\,$\pm$\,2.3)\,$\times$\,$10^{22}$\un{cm^{-2}}.
The bremsstrahlung column density was found to be 11.3\,$\pm$\,2.5 with a 
temperature around 15\un{keV} but once again the error range is so large that
this value cannot be constrained.
The low statistics severly limit our ability to detect spectral variations during the flares.
The average 2--10\un{keV} luminosity of this flare is around
4.3\,$\times$\,$10^{34}$\ergpersec (see Table\,\ref{t:specParams}) and as in the
case of the March 31 flare, we estimate the peak luminosity to be about twice the 
average, and thus reached a maximum intensity around 40 times 
the quiescent luminosity of \sgra.

A summary of the basic temporal features of the two factor-40 flares are given
in Table\,\ref{t:flareBasics}. This includes the start, peak and end times of the 
flares given in UTC and \xmm time in seconds to facilitate reference to the data.
We also listed the peak count rate and the deviation from the mean in units of sigma.

\begin{deluxetable}{lcccc}
\tabletypesize{\scriptsize}
\setlength{\tabcolsep}{2mm}
\tablewidth{0pt}
\tablecolumns{3}
\tablecaption{Basic features of \xray flares}
\tablehead{
\colhead{}			& \multicolumn{2}{c}{2004 March 31} & \multicolumn{2}{c}{2004 August 31} \\
\colhead{}			& \colhead{UTC} & \colhead{XMM time (s)} & \colhead{UTC} & \colhead{XMM time (s)}  }

\startdata
Start time		\dotfill & 22:48 &	1.9716050+08	& 09:58 &	2.1033350+08	\\
Peak time		\dotfill & 23:09 &	1.9716175+08	& 11:05 &	2.1033750+08	\\
End time 		\dotfill & 23:22 &	1.9716250+08	& 11:22 &	2.1033850+08	\\
Duration (s)		\dotfill & \multicolumn{2}{c}{2\,500} & \multicolumn{2}{c}{5\,000} \\
Peak rate (cts/s)	\dotfill & \multicolumn{2}{c}{$0.760\pm0.048$} & \multicolumn{2}{c}{$0.617\pm0.035$} \\
Mean rate (cts/s)	\dotfill & \multicolumn{2}{c}{$0.303\pm0.002$} & \multicolumn{2}{c}{$0.293\pm 0.002$} \\
Deviation  ($\sigma$)\dotfill & \multicolumn{2}{c}{9.6}		& \multicolumn{2}{c}{9.7}	
\enddata
\tablecomments{\scriptsize Error bars correspond to the 68$\%$ confidence interval. 
	Count rates and times are estimated using 500\un{s} time bins. } 

\label{t:flareBasics}
\end{deluxetable}

\subsection{Timing Analysis}

The cyclical decrease in flux observed during ObsID\,866 
(first portion of Figure\,\ref{f:lc_866-867}) occurs with a period of about 8\un{h} 
and is most probably due to an eclipse in the binary system \cxo. 
A complete analysis of the \xmm obsevations of this transient source are 
presented by Porquet et al. (2005), and the results of the \chandra 
observations by Muno et al. (2005b).

A spectral density analysis of the \xray data from \sgra was 
performed and since \xmm's spatial resolution severely limits our ability
to distinguish the quiescent emission of \sgra from that of its
surroundings, temporal structures in the \xrays emission detected
by \xmm from the Galactic center can only be attributed to \sgra
with confidence if seen during a flare, when the flux rises 
substantially above the quiescent emission.
Therefore, periodic features seen in the base level of the light curve
cannot readily be attributed to \sgra.
Furthermore, a period detected during a flare can only be confidently 
attributed to \sgra if it is not present in the rest of the light curve.

Since the range of frequencies and resolution in a periodogram
are primarily functions of the total length of the observation and 
sampling or bin time, only the second of the two factor-40 flares
allowed us to perform a meaningful search for periodic modulations.
This event lasted about 10\un{ks} including the precursor. 
This is about 4 times longer than the first. 

The first analysis revealed a periodic structure that could be significant 
but the proper treatment of the signal that includes white and 
possibly a component of red noise, requires a detailed analysis that is quite 
complex and that we have begun but will present elsewhere upon completion.

\section{Discussion and Conclusion}
\label{s:discussion}

We observed the neighborhood of \sgra for more than three consecutive days 
from 2004 March 28 to April 1, and an additional three consecutive days 
from 2004 August 31 to September 3.  Both observations were interrupted 
for only $\approx 30$\un{ks} and during each of these observation periods 
we detected flaring activity composed in each case of a large flare
accompanied by smaller ones.  The two main flares are positionally coincident 
with \sgra and, as Baganoff et al.\ (2001), we associate
these with the supermassive black hole at the center of the Milky Way, 
given our current knowledge of the region surrounding this source. 
These flares, both with peak luminosities about 40 times that of
the assumed quiescent level, exhibit similar spectral characteristics:
a hard photon index of $\Gamma$\,$\approx$\,1.5--1.9 and a
column density $N_{\rm H}$\,$\approx$\,8--13\,$\times$\,$10^{22}$\un{cm^{-2}}.
These parameter values are more akin to those of the first two detected flares 
(Baganoff et al.\ 2001; Goldwurm et al.\ 2003),
than to those of the very bright flare reported by Porquet et al.\ (2003).

There were a total of three flares with significance greater than
3$\sigma$, corresponding to a factor of 15 above \sgra's quiescent level 
in ObsID\,789 and the same number in ObsID\,866,
if we consider the precursor to be a separate flaring event.
Although we cannot determine the position of the smaller flares
accurately and thus cannot attribute them to \sgra with certainty,
if these are indeed coming from the central black hole then,
irrespective of the temporal distribution of these flares, 
our estimate of their average occurrence rate is about 1 per day.
This estimate agrees with those based on \chandra observations performed
between 1999 and 2002 \citep{c:baganoff03b}.
Note that flares from \sgra appear to occur in clusters.

In light of the recently detected near-IR spectrum of \sgra \citep{c:genzel03}, 
and of the constraints that it imposes on the various emission models for 
this source, the flares we detected could have resulted from a sudden
increase in accretion accompanied by a reduction in the anomalous viscosity,
which would point to thermal bremsstrahlung as the \xray emission process.
This would give rise to a hard spectrum, unlike the softer one expected 
from a synchrotron self-Compton process \citep{c:liu02}.

Another possibility is that the flares arose from the quick acceleration 
of electrons in the accretion flow near the black hole, producing either
a two-component distribution characterized by a broken power-law, with 
steep low-energy and flat high-energy components, or a modified thermal 
distribution with a high-energy enhancement (see, e.g. Liu, Petrosian,
\& Melia 2004; Yuan, Quataert, \& Narayan\ 2003 and 2004).  
In either of these cases, the lower energy electrons 
produce the near-IR emission and the high-energy distribution produces the 
hard \xrays.  Further, the rather short time scale ($\sim$\,2500--4500\un{s}) 
and hard spectral index ($\Gamma$\,=\,1.6) of the flares would,
according to this model, favor magnetic reconnection as the engine of the event.

In either of these two scenarios --- an accretion instability, or a magnetic
reconnection event --- one could expect to see a modulation in the 
light curve, mirroring the underlying Keplerian period of the emitting plasma.
Indications of the presence of such a periodic modulation during a near-IR flare
were presented by Genzel et al.\ (2004) and during an \xray flare 
by Ascenbach et al.\ (2002). 
These results must be confirmed by similar detections in other events
seen in the near-IR and \xray bands.
A search for such a semi-periodic modulation in the longer of the two factor-40
flares from \sgra presented here is under way and the results will be reported elsewhere. 

The accretion instability scenario predicts a strong correlation 
between the sub-mm/IR and the \xray photons with no or little change in 
the millimeter flux density.  The two-component synchrotron model predicts possible,
though not necessary, correlations between \xray and near-IR flares, with larger \xray 
amplitudes in the case of simultaneous flaring, important variations in the spectral 
slopes of the \xray flares compared with those in the near-IR flares, and small amplitude 
variability in the radio and sub-millimeter wavebands.  It is on this last point that 
we could distinguish the accretion induced \xray flare model from the two-component 
synchrotron model.

\acknowledgements{
G.\;B\'elanger would like to thank Jean Ballet, Monique Arnaud, Jean-Luc Sauvageau
and Anne Decourchelle for their kind help with several aspects of the \xmm analysis, 
and Michael Muno, R\'egis Terrier and Matthieu Renaud for several useful discussions. 
G.\;B. is grateful to the referee whose comments were very pertinent and lead to a 
finer analysis and deeper understanding of the statistical methods used in this task.
G.\;B. acknowledges the financial support from the French Space Angency (CNES).
This work is based on observations obtained with \xmm,
an ESA science mission with instruments and contributions 
directly funded by ESA member states and the USA (NASA).
}

\clearpage

\begin{turnpage}
\begin{deluxetable}{ccccccccccccc}

\tabletypesize{\footnotesize}
\setlength{\tabcolsep}{0mm}
\tablewidth{0pt}
\tablecolumns{13}
\tablecaption{Spectral Characteristics of \xray Flares}
\tablehead{
\colhead{} & \multicolumn{5}{c}{Power-law} & \colhead{} & \multicolumn{6}{c}{Black Body} \\
\cline{2-6} \cline{8-13} \\
\colhead{} & \colhead{} & \colhead{} & \colhead{$F_{\rm{X}}$[2--10]} & \colhead{} & \colhead{} &
\colhead{} & \colhead{} & \colhead{} & \colhead{} & \colhead{$F_{\rm{X}}$[2--10]} & \colhead{} & \colhead{} \\
\colhead{} & \colhead{$N_{\rm{H}}$} & \colhead{} & \colhead{($10^{-12}$\,erg} & \colhead{$L_{\rm{X}}$[2--10]} & \colhead{} & 
\colhead{} & \colhead{$N_{\rm{H}}$} & \colhead{kT} & \colhead{Norm} & \colhead{($10^{-12}$\,erg} & \colhead{$L_{\rm{X}}$[2--10]} & \colhead{} \\
\colhead{Date (Instr.)} & \colhead{($10^{22}$\un{cm^{-2}})} & \colhead{$\Gamma$} & \colhead{\percmsec)} & \colhead{($10^{34}$\ergpersec)} & \colhead{C/bins} & 
\colhead{} & \colhead{($10^{22}$\un{cm^{-2}})} & \colhead{(keV)} & \colhead{($10^{-5}$)} & \colhead{\percmsec)} & \colhead{($10^{34}$\ergpersec)} & \colhead{C/bins}
}

\startdata
March\,31$\;$(PN) & $8.0^{+3.3}_{-2.8}$ & $1.7^{+0.6}_{-0.6}$ & $4.0^{+0.4}_{-0.9}$ & $5.1^{+0.3}_{-0.7}$ & 689/740 & & 
		$3.4^{+2.3}_{-1.9}$ & $1.9^{+0.5}_{-0.3}$  & $7.4^{+1.6}_{-1.2}$ & $4.0^{+1.2}_{-1.6}$ & $3.6^{+0.7}_{-1.4}$ & 691/740 \\ \\
March\,31$\;$(All) & $8.3^{+2.5}_{-2.1}$ & $1.5^{+0.5}_{-0.4}$ & $4.0^{+0.3}_{-0.4}$ & $4.9^{+0.9}_{-0.7}$ & 883/936 & &
		     $4.1^{+1.7}_{-1.5}$ & $1.9^{+0.2}_{-0.3}$ & $7.4^{+1.6}_{-0.8}$ & $3.8^{+0.7}_{-0.9}$ & $4.6^{+0.5}_{-0.9}$ & 886/936 \\ \\

\cline{2-13} \\

August\,31$\;$(PN) & $10.0^{+3.4}_{-2.5}$ & $1.3^{+0.5}_{-0.5}$ & $2.7^{+0.2}_{-0.4}$ & $3.4^{+0.8}_{-0.5}$ & 883/731 & &
		$6.5^{+2.3}_{-1.7}$ & $2.1^{+0.4}_{-0.3}$ & $5.9^{+1.5}_{-0.8}$ & $2.6^{+0.5}_{-0.9}$ & $2.6^{+0.4}_{-0.8}$ & 883/731 \\ \\
August\,31$\;$(All) & $12.5^{+3.4}_{-2.8}$ & $1.9^{+0.5}_{-0.5}$ & $2.7^{+0.2}_{-0.4}$ & $4.3^{+1.2}_{-0.8}$ & 1082/913 & &
		     $7.1^{+2.3}_{-2.1}$ & $1.8^{+0.3}_{-0.2}$ & $5.7^{+0.9}_{-0.5}$ & $2.6^{+0.4}_{-0.4}$ & $2.9^{+0.3}_{-0.7}$ & 1086/913 \\ \\
\enddata
\tablecomments{\footnotesize Errors on fitted parameters correspond to the 68$\%$ confidence interval.
		Flux density for the Power-law model is normalized over the range from 
		2 to 10 keV and corresponds to the total unabsorbed flux in this range. 
		The listed flux values are absorbed with the associated column density and
		the luminosity is calculated using the power-law normalization and is 
		derived for a distance of 8\un{kpc} to the Galactic center.	
		In the Black Body model, the normalization corresponds to the dimensionless 
		ratio of the luminosity in units of $10^{39}$\ergpersec to the distance in units 
		of 10\un{kpc} squared. In the first column, 'Instr.' refers to the instrument whose 
		data was used in the fit and 'All' refers to the combination of PN, Mos\,1 and Mos\,2.} 

\label{t:specParams}
\end{deluxetable}
\end{turnpage}

\end{document}